\documentclass[12pt,epsf,epsfig,psfig]{article}
\usepackage{graphicx}
\usepackage{epsfig}
\usepackage{cite}
\def\be{\begin{equation}}
\def\ee{\end{equation}}

\def\lsim{\raise0.3ex\hbox{$<$\kern-0.75em\raise-1.1ex\hbox{$\sim$}}}
\def\gsim{\raise0.3ex\hbox{$>$\kern-0.75em\raise-1.1ex\hbox{$\sim$}}}

\def\NP{{ Nucl.\ Phys.\ }}

\def\PR{{ Phys.\ Rev.\ }}

\def\EP{{ Europ.\ Phys.\ J.\ C}}

\begin{document}
\centerline{\bf High Energy Hadron Production, Self-Organized Criticality}
\centerline{\bf  and Absorbing State Phase Transition}

\vskip1cm

\centerline{\bf Paolo Castorina$^{a,b}$ and Helmut Satz$^c$}

\centerline{a: INFN sezione di Catania, Catania, Italy}

\centerline{b: Faculty of Physics and Mathematics, Charles University, Prague,
Czech Republic}

\centerline{c: Fakult\"at f\"ur Physik, Universit\"at Bielefeld, Germany}

\vskip1.0cm

\begin{abstract}
In high energy nuclear collisions, production rates of light nuclei agree with the 
predictions of an ideal gas at a temperature $T=155 \pm 10$ MeV. In an 
equilibrium hadronic medium of this temperature, light nuclei cannot survive. 
In this contribution, we suggest that the observed behavior is due to an evolution in global 
non-equilibrium, leading to self-organized criticality and to hadron formation 
as an absorbing state phase transition for color degrees of freedom.
At the confinement  point, the initial quark-gluon medium becomes quenched by the vacuum, breaking 
up into all allowed free hadronic and nuclear mass states, without  (or with a very short-live)
 subsequent formation of thermal hadronic medium.
\end{abstract}

\section{Introduction}
The yields for  deuteron, ${^3} He$, hyper-triton, ${^4} He$ and
their antiparticles  have recently been measured  in $Pb-Pb$
collisions by the ALICE collaboration \cite{Alice1,Alice2,Alice3} and are in very good agreement \cite{andronic}
with the statistical hadronization model (SHM) \cite{shm},  with a formation temperature of $T \simeq, 155$ MeV, corresponding 
at the (pseudo)critical confinement 
temperature $T_c = 155 \pm 10$ MeV \cite{baza1}. 

The curious feature is that the all hadron abundances are already specified 
once and for all at $T_c$ and are not subsequently modified in the evolution
of the hadron gas and this  enigma is further enhanced by the yields
for light nuclei. Indeed, these states have binding energies of a few MeV and are generally
much larger than hadronic size, and therefore their survival in the assumed hot hadron gas
poses an even more striking puzzle \cite{pbm}. For example, the hyper-triton root-mean-square size
is close to $10$ fm, about the same size of the whole fireball formed in Pb-Pb collision at $\sqrt s = 2.76$ TeV, and the energy needed to remove the $\Lambda$ from it 
is  $130 \pm 30$ KeV.

Why are the yields for the production of light nuclei determined by the 
rates as specified at the critical hadronization temperature, although in hot 
hadron gas they would immediately be destroyed?

In this contribution, following ref. \cite{noi}, we want to discuss a solution to this puzzle
obtained by abandoning the idea of a thermal hadron medium existing below
the confinement point. 
We propose that the hot quark-gluon system, 
when it cools down to the hadronization temperature, is effectively quenched 
by the cold physical vacuum. 

The relevant basic mechanism for this is 
self-organized criticality, leading to universal scale-free behavior,
based on an absorbing state phase transition for color degrees of freedom.

\section{Self-organized criticality and absorbing state phase transition}

The core hypothesis of Self-Organized Criticality (SOC) \cite{bak,lib} is  that systems consisting of many interacting components will, under certain conditions, spontaneously organize into a state with properties akin to that ones observed in a equilibrium thermodynamic system, as  the  scale-free behavior. 
 
The self-organized evolution indicates that the complex behavior arises spontaneously without the need for the external tuning of a control parameter  (the temperature for example). In SOC the dynamics of the order parameter drives the control parameter to the critical value:  natural dynamics drives the system towards and  maintains it at the edge of stability.

For non-equilibrium steady states it is becoming increasingly evident 
that  SOC is related to conventional critical behavior by the concept of absorbing-state phase transition \cite{hh1,hh2}.

An absorbing state is a configuration that can be reached by the dynamics but cannot be left  and
absorbing state phase transitions are among the simplest non-equilibrium phenomena displaying critical behavior and universality \cite{hh1,hh2}.

A clear example is given by models describing the growth of bacterial colonies or the
spreading of an infectious disease among a population: once an absorbing state, e.g., a state in which all the bacteria are dead, is reached, the system cannot escape from it.

Let us now consider the hadronization dynamics where, for sake of simplicity, an initial $e^+e^-$ annihilation produces a $\bar q q$ pair which evolves according to QCD dynamics.
The short distances dynamics is due to local interacting color charges , with the QCD processes of parton (quarks and gluons)  annihilations and creation.

The dynamics of color degrees of freedom (d.o.f.) ends up with the hadronic production, i.e. with the production of colorless clusters. The final state has no color and the evolution of the system cannot produce colored partons in the final state.

From this point of view, hadron production is a phase transition to an absorbing state for color degrees of freedom.
Moreover this phase transition is a non-equilibrium one, since, by definition, the rate out of an absorbing state is zero and an absorbing state can not obey the detailed balance.

A toy model which shows how the competition between hadron (h) formation, i.e. color neutralization, and production and/or annihilation of color charges (partons $P$) leads to an absorbing state is easily obtained by considering a normalized quantity  $\rho(t)$, proportional to color charge,, as a function of time $t$ and the processes: $P+P \rightarrow P$ with rate $\lambda$ (parton annihilation); $P \rightarrow P+P$ with rate $\sigma$ (parton production); $P \rightarrow h$ with rate $k$ (color neutralization).

The mean field evolution equation is given by \cite{hh1,hh2}
\be
\frac{d \rho}{dt} = (\sigma - k) \rho -\lambda \rho^2 = \rho( \sigma - k - \lambda \rho) \,\,\, .
\ee

If $\sigma < k$ the steady state is $\rho_s=0$ and  is an absorbing state. If $\sigma > k$ the steady state is $\rho_s = (\sigma - k) / \lambda$, the critical value is $\sigma_c=k$ and, as in thermal equilibrium, the critical point is governed by a power law behavior $ \rho_s \simeq (\sigma -\sigma_c)^\beta$ with $\beta = 1$. 

Absorbing states characterize first order phase transitions also \cite{altro} and, indeed, for pure $SU(N)$ gauge theories, where the Polyakov loop, $l$, is an order parameter, one can show that the dynamical evolution of the system \cite{pis} has a steady state with $l=0$, which is an absorbing state.

According to previous discussion:  1) The Hadronization mechanism is  a non equilibrium phase transition  to an absorbing  state for color d.o.f.; 2) The dynamical evolution is driven by
color d.o.f.  up to the hadronization time/temperature; 3) a natural assumption, due to the absorbing state phase transition, is that the system 
is essentially «frozen» at the values of the  parameters at  the transition.

Let us discuss the consequences of this point of view for the hadron production.

\section{SOC in hadron formation}

\subsection{Self-organization and hadronic spectrum }

The typical illustration of SOC, proposed in the pioneering work \cite{bak}, 
is the avalanches dynamics of sandpiles, where the number $N(s)$ of  avalanches of size $s$ 
observed over a long period is found to vary as a power of $s$, $N(s) = \alpha s^{-p}$, which means 
that the phenomenon is scale-free. 

Another useful example of self-organized criticality provided by 
partitioning integers \cite{noi}. Consider the {\sl ordered} partitioning of an integer 
$n$ into integers. The number $q(n)$ of such partitionings is for $n=3$
equal to four: 3, 2+1, 1+2, 1+1+1, i.e., $q(3)=4$. It is easily shown 
\cite{blan} that in general $
q(n) = 2^{n-1}$, i.e. the
number of partitions increases exponentially with the size of the integer.
 
Given an initial integer $n$, we would now like to know the number $N(k,n)$
specifying how often a given integer $k$ occurs in the set of all 
partitionings of $n$. To illustrate, in the above case of $n=3$, we have 
$N(3,3)=1$, $N(3,2)=2$ and $N(3,1)=5$. To apply the formalism of 
self-organized criticality, we have to attribute a strength $s(k)$ to each
integer. It seems natural use the number of partitions for this, i.e., set
$s(k) = q(k)$ and the desired number $N(k,n)$ in a scale-free scenario is then given by
\be
N(k,n) = \alpha(n) [s(k)]^{-p}.
\label{p3}
\ee
For small values of $n$, $N(k,n)$ is readily
obtained explicitly and one finds  that the critical
exponent becomes $p \simeq 1.26$. 

The previous example is 
immediately reminiscent of the statistical bootstrap model
of Hagedorn \cite{hagedorn}, who had ``fireballs composed of fireballs, which
in turn are composed of fireballs, and so on''. Indeed, its general pattern
has been shown to be due to an underlying structure analogous to the partitioning of an integer into integers 
\cite{blan}.  

More precisely, Hagedorn's bootstrap approach \cite{hagedorn} proposes that a hadronic colorless
state of overall mass $m$ can be partitioned into structurally similar
colorless states, and so on. If these states were at rest, the situation would be
identical to the above partioning problem. Since the constituent fireballs 
have an intrinsic motion, the number of states $\rho(m)$
corresponding to a given
mass $m$ is determined by the bootstrap equation which can be asympotically  solved  \cite{nahm}, giving
$\rho(m) \sim m^{-a}$ exp${(m/T_H)}$
and $T_H$ as solution of 
\be
\left({2\over 3 \pi}\right)\left(T_H \over m_0\right) 
K_2(m_0/T_H) = 2 \ln 2 - 1,
\label{bs}
\ee
with $m_0$ denoting the lowest possible mass and $K_2(x)$ is a Hankel function of pure imaginary argument. For
$m_0 = m_{\pi} \simeq 130$ Mev, this leads to the Hagedorn temperature
$T_H \simeq 150$ MeV, i.e., to approximately the critical hadronization 
temperature found in statistical QCD. The cited solution gave $a=3$,
but other exponents could also been considered. 

The  previous form  is an asymptotic solution of the bootstrap
equation which  diverges for $m\to 0$ and must be modified for 
small masses. Using a similar result 
for $\rho(m)$ obtained in the dual resonance model \cite{huang-wein}, 
Hagedorn proposed  
\be
\rho(m) = {\rm const.}(1+ (m/\mu_0))^{-a} \exp(m/T_H) 
\label{exp1}
\ee   
where $\mu_0 \simeq 1 - 2$ GeV is a normalization constant.

At this point we should emphasize that the form of $\rho(m)$ is entirely due to the self-organized nature of the 
components and they are in no way a result of thermal behavior. We have expressed
the slope coefficient of $m$  in terms of the Hagedorn 
``temperature'' only in reference to subsequent applications. In itself, it
is totally of combinatorical origin.

\subsection{Comparison with ALICE data}

We now  apply the formalism of self-organized criticality to strong
interaction physics.  Our picture assumes a sudden quench of the partonic medium produced
in the collision. The initial hot system of deconfined quarks and 
gluons rapidly expands and cools; while this system is presumably in local
thermal equilibrium, the difference between transverse and longitudinal
motion implies a global non-equilibrium behavior. The longitudinal
expansion quickly drives the system to the hadronisation point, and it is 
now suddenly thrown into the cold physical vacuum. The process is not 
unlike that of a molten metal being dumped into cold water. In this quenching
process, the system freezes out into the degrees of freedom presented by 
the system at the transition point and subsequently remains as such, due to the absorbing state nature of the transition, apart
from possible hadron or resonance decays. There
never is an evolving warm metal. In other words, in our case there is no hot 
 (or a very short-live) interacting hadron gas.  Whatever
thermal features are observed, such as radial or elliptic hydrodynamic
flow, must then have originated from local equilibrium in the earlier 
deconfined stage \cite{localSOC1}.
The mechanism driving the system rapidly to the critical point
is the global non-equilibrium due to the longitudinal motion provided
by the collision.

In such a scenario, high energy nuclear collisions lead to a system 
which at the critical point, i.e. the color absorbing state, breaks up into
components of different masses $m$, subject to self-similar
composition and hence of a strength $\rho(m)$
as given by the above Eq.\ (\ref{exp1}). In the self-organized criticality
formalism, this implies that the interaction will produce
\be
N(m) = \alpha [\rho(m)]^{-p}
\label{p5}
\ee
hadrons of mass $m$. With $\rho(m)$ given by Eq.\ (\ref{exp1}), the 
resulting powerlaw form
\be
\log N(m) = -m \left({p \log e \over T_H} \right) \left[1 
- \left(a T_H \over
m \right ) 
{\ln(1+{m \over \mu_0})}\right]
+ {\rm const.} 
\label{p6}
\ee
is found to show a behavior similar to that obtained from 
an ideal resonance gas in equilibrium. We emphasize that it 
is here obtained assuming only scale-free
behavior (self-organized criticality) and a mass weight determined by
the number of partitions. No equilibrium thermal system of any kind is assumed.

\begin{figure}[htb]
\centerline{\psfig{file=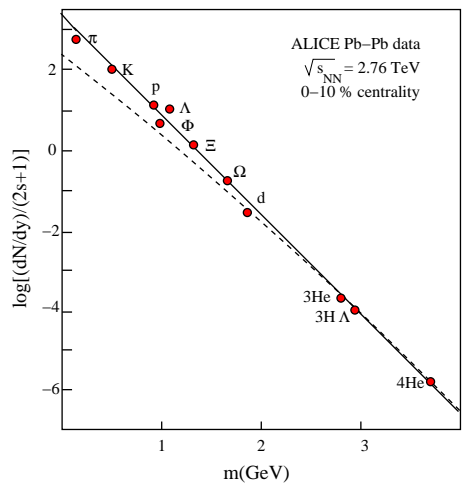,width=6cm}} 
\caption{Yield rates of species at central 
rapidity vs. their mass $m$ \cite{Alice1,Alice2,Alice3}. The solid line corresponds to
Eq.\ (\ref{p7}), the dashed line to Eq.\ (\ref{p7a}).}
\label{soc3}
\end{figure}

We now consider the mentioned ALICE data \cite{Alice1,Alice2,Alice3}. In Fig.\ \ref{soc3}
the production yields for the different mass states in central
$Pb-Pb$ collisions at $\sqrt s = 2.76$ GeV are shown; in each 
case, the yield is divided by the relevant spin degeneracy. We see that the
yields show essentially powerlike behavior, and the light nuclei follow the 
same law as the elementary hadrons. The solid line 
in Fig.\ \ref{soc3} shows the behavior obtained from eqs.\ 
(\ref{p6}), ignoring for the moment the second term in the square brackets,
\be
\log[(dN/dy)/(2s+1)] \simeq -m\left({0.43~\!p\over T_H}\right) + A,
\label{p7}
 \ee
with $T_H=0.155 MeV$ and fit values $p=0.9$, $A=3.4$. 
The form is evidently in good agreement with the data.

\medskip

Including the correction term to linear behavior that we had omitted above, 
we have
\be
\log[(dN/dy)/(2s+1)] \simeq -m\left({0.43~\!p\over T_H}\right) + 
+ p~\!a\log[1+(m/\mu)] + A.
\label{p7a}
 \ee
The additional term is, as indicated, rather model dependent. It will 
effectively turn the yield curve down for decreasing masses. This is in fact 
necessary, since the decay of heavier resonances will enhance the direct 
low mass meson yields.
To illustrate the effect of the term, we choose $a=3$, corresponding to
the mentioned solution of the bootstrap equation \cite{nahm},
and $\mu=2$ GeV for the normalization. The result is included in Fig.\
\ref{soc3}.

\medskip 
 {\bf Acknowledgements}
P.C. thanks the School of Nuclear Science and Technology, Lanzhou University, for the hospitality and Marco Ruggieri and Pengming Zhang for useful comments.

\medskip 

\end{document}